\pdfoutput=1
\documentclass[preprint,12pt]{elsarticle}
\usepackage{graphicx}

\usepackage{amsmath, amssymb}

\biboptions{sort&compress, square, comma}

\usepackage[linkcolor=blue, citecolor=blue, colorlinks=true]{hyperref}
\usepackage{booktabs}

\usepackage{multirow}

\usepackage{algorithm}
\usepackage{algorithmic}

\providecommand{\e}[1]{\ensuremath{\times 10^{#1}}}

\journal{Combustion and Flame}

\begin{document}

\begin{frontmatter}


\title{On the importance of graph search algorithms for DRGEP-based mechanism reduction methods}

\author[cwru]{Kyle~E.\ Niemeyer}
%\ead{ken7@case.edu}

\author[uconn]{Chih-Jen Sung\corref{cor1}}
\ead{cjsung@engr.uconn.edu}

\address[cwru]{Department of Mechanical and Aerospace Engineering\\
	Case Western Reserve University, Cleveland, OH 44106, USA}
\address[uconn]{Department of Mechanical Engineering\\
	University of Connecticut, Storrs, CT 06269, USA}
	
\cortext[cor1]{Corresponding author}

\begin{abstract}
The importance of graph search algorithm choice to the directed relation graph with error propagation (DRGEP) method is studied by comparing basic and modified depth-first search, basic and \emph{R}-value-based breadth-first search (RBFS), and Dijkstra's algorithm.  By using each algorithm with DRGEP to produce skeletal mechanisms from a detailed mechanism for \emph{n}-heptane with randomly-shuffled species order, it is demonstrated that only Dijkstra's algorithm and RBFS produce results independent of species order.  In addition, each algorithm is used with DRGEP to generate skeletal mechanisms for \emph{n}-heptane covering a comprehensive range of autoignition conditions for pressure, temperature, and equivalence ratio.  Dijkstra's algorithm combined with a coefficient scaling approach is demonstrated to produce the most compact skeletal mechanism with a similar performance compared to larger skeletal mechanisms resulting from the other algorithms.  The computational efficiency of each algorithm is also compared by applying the DRGEP method with each search algorithm on the large detailed mechanism for \emph{n}-alkanes covering \emph{n}-octane to \emph{n}-hexadecane with 2115 species and 8157 reactions.  Dijkstra's algorithm implemented with a binary heap priority queue is demonstrated as the most efficient method, with a CPU cost two orders of magnitude less than the other search algorithms.
\end{abstract}


\end{frontmatter}

%\pagebreak


\section{Introduction}
\label{S:intro}

The directed relation graph (DRG) method for the skeletal reduction of large reaction mechanisms, originally developed by Lu and Law \cite{Lu:2005,Lu:2006}, has been shown to be applicable to relevant transportation fuel components \cite{Lu:2006a,Lu:2008}.  Development of this approach has since focused on variants of the method \cite{Zheng:2007, Pepiot-Desjardins:2008, Sun:2010, Niemeyer:2010}, but DRG with error propagation (DRGEP) in particular has received much attention \cite{Pepiot-Desjardins:2008,Zsely:2009,Liang:2009,Liang:2009a,Niemeyer:2010,Shi:2010,Shi:2010a}.  The DRGEP differs mainly from DRG in that it takes error propagation down graph pathways into account.  Further improvement of the DRGEP method is motivated by the large and continually increasing size of detailed reaction mechanisms for liquid transportation fuels \cite{Lu:2009}; for instance, see the recent biodiesel surrogate mechanism of Herbinet et al.\ \cite{Herbinet:2010} that contains 3299 species and 10806 reactions.

The DRG method maps the coupling of species onto a directed graph and finds unimportant species for removal by eliminating weak graph edges using an error threshold, where the graph nodes represent species and graph edge weights indicate the dependence of one species on another.  Following a simple graph search initiated at certain preselected target species (e.g.,\ fuel, oxidizer, important pollutants), species unreached by the search are considered unimportant and removed from the skeletal mechanism.  DRGEP modifies this approach slightly by considering the dependence of a target on other species due to a certain path as the product of intermediate edge weights and the overall dependence is the maximum of all path dependencies.

One point that has not received much attention is the choice of graph search algorithm used to calculate this overall dependence.  While the DRG method only needs to find connected graph nodes and can therefore use any search algorithm, caution must be taken when selecting the method used with DRGEP.  The issue of calculating the overall dependence is actually a single-source shortest path problem \cite{Cormen:2001} where the ``distance'' of the path is the product of intermediate edge weights rather than the sum, and the ``shortest'' path is that which has the maximum overall dependence rather than the minimum.  Search algorithms that in general do not correctly calculate the overall dependence will underestimate the importance of species and cause premature removal from the skeletal mechanism.  Further, the resulting skeletal mechanism is independent of the order of the species in the mechanism only if a specific algorithm can correctly capture and calculate the overall dependence of species.  Therefore, the reliability of various algorithms needs to be studied by determining whether each is dependent on the order of species in a detailed mechanism.

The efficiency of the graph search algorithm is also an important factor due to the recent use of DRGEP in dynamic skeletal reduction approaches \cite{Liang:2009,Liang:2009a,Shi:2010}.  In the worst case, DRGEP is applied at every spatial grid location and each time step to generate locally relevant skeletal mechanisms, although as Shi et al.\ \cite{Shi:2010} demonstrated this can be eased by combining dynamic DRGEP-based reduction with an adaptive multi-grid chemistry model.  Regardless, the computational cost of the search algorithm increases with the number of species and must be considered when comparing algorithms due to the large and ever-increasing size of detailed reaction mechanisms.

Most DRGEP studies reported using algorithms based on either depth-first search (DFS) \cite{Niemeyer:2010} or breadth-first search (BFS) \cite{Liang:2009,Liang:2009a,Shi:2010,Shi:2010a}, but no comparison has been performed.  Here we compare such methods with Dijkstra's algorithm, the classical solution to the single-source shortest path problem \cite{Dijkstra:1959,Cormen:2001}, in order to demonstrate the weaknesses of DFS- and BFS-based approaches and the subsequent effectiveness and reliability of Dijkstra's algorithm.  First, by randomly shuffling the order of species in the detailed mechanism for \emph{n}-heptane of Curran et al.\ \cite{Curran:1998}, we show that DFS- and BFS-based algorithms can generate results dependent on the order of species in the reaction mechanism while Dijkstra's algorithm generates consistent results regardless of species order.  Second, we demonstrate that, for a given error limit, more compact skeletal mechanisms are possible with DRGEP by using Dijkstra's algorithm.  This is done by comparing skeletal mechanisms generated with different search algorithms for \emph{n}-heptane covering a comprehensive range of autoignition conditions for pressure, temperature, and equivalence ratio.  Third, we compare the computational efficiency of the various search algorithms by calculating the CPU time required to perform the graph search on the large detailed mechanism of Westbrook et al.\ \cite{Westbrook:2009} covering oxidation of \emph{n}-alkanes from \emph{n}-octane to \emph{n}-hexadecane containing 2115 species and 8157 reactions.

\section{Methodology}
\label{S:method}

\subsection{DRGEP Method}

In the current work we use the DRGEP method of Pepiot-Desjardins and Pitsch \cite{Pepiot-Desjardins:2008}, described here in brief.  Accurate calculation of the production of a species \emph{A} that is strongly dependent on another species \emph{B} requires the presence of species \emph{B} in the reaction mechanism.  This dependence is expressed with the direct interaction coefficient (DIC):
\begin{equation}
	r_{AB} = \frac{ \bigl | \sum_{i = 1}^{n_R} \nu_{A, i} \omega_{i} \delta_{B}^{i} \bigr | } {\max \left( P_A, C_A \right)  },		\label{E:rAB}
\end{equation}
where
\begin{align}
	P_A &= \sum_{i=1}^{n_R} \max \left( 0, \nu_{A,i} \omega_i \right),		\label{E:PA}\\
	C_A &= \sum_{i=1}^{n_R} \max \left( 0, - \nu_{A,i} \omega_i \right),		\label{E:CA}
\end{align}
\begin{equation}
	\delta_{B}^{i} = \begin{cases}
		1 &\text{if reaction \emph{i} involves species \emph{B},} \\
		0 &\text{otherwise,} \end{cases}							\label{E:dBi}
\end{equation}
\emph{A} and \emph{B} represent the species of interest (with dependency in the $A \rightarrow B$ direction meaning that \emph{A} depends on \emph{B}), \emph{i} the \emph{i}th reaction, $\nu_{A, i}$ the stoichiometric coefficient of species \emph{A} in the \emph{i}th reaction, $\omega_i$ the overall reaction rate of the \emph{i}th reaction, and $n_R$ the total number of reactions.

After calculating the DIC for all species pairs, a graph search is performed starting at user-selected target species to find the dependency paths for all species from the targets.  A path-dependent interaction coefficient (PIC) represents the error propagation through a certain path and is defined as the product of intermediate DICs between the target \emph{T} and species \emph{B} through pathway \emph{p}:
\begin{equation} \label{E:rABp}
	r_{T B, p} = \prod_{j = 1}^{n-1} r_{S_j S_{j + 1}}	,
\end{equation}
where \emph{n} is the number of species between \emph{T} and \emph{B} in pathway \emph{p} and $S_j$ is a placeholder for the intermediate species \emph{j} starting at \emph{T} and ending at \emph{B}.  An overall interaction coefficient (OIC) is then defined as the maximum of all PICs between the target and each species of interest:
\begin{equation} \label{E:RAB}
	R_{T B} = \max_{\text{all paths $p$}} \left( r_{T B, p} \right)	.
\end{equation}

Pepiot-Desjardins and Pitsch \cite{Pepiot-Desjardins:2008} also proposed a coefficient scaling procedure to better relate the OICs from different points in the reaction system evolution that we adopt here.  A pseudo-production rate of a chemical element \emph{a} based on the production rates of species containing \emph{a} is defined as
\begin{equation}	\label{E:Patom}
	P_a = \sum_{\text{all species } S} N_{a, S} \max \left( 0, P_S - C_S \right) ,
\end{equation}
where $N_{a, S}$ is the number of atoms \emph{a} in species \emph{S} and $P_S$ and $C_S$ are the production and consumption rates of species \emph{S} as given by Eqs.\ \ref{E:PA} and \ref{E:CA}, respectively.  The scaling coefficient for element \emph{a} and target species \emph{T} at time \emph{t} is defined as
\begin{equation}	\label{E:alpha_aT}
	\alpha_{a, T} (t) = \frac{ N_{a, T} \left| P_T - C_T \right| }{ P_a } .
\end{equation}
For the set of elements $ \lbrace \mathcal{E} \rbrace $, the global normalized scaling coefficient for target \emph{T} at time \emph{t} is
\begin{equation}	\label{E:alpha_T}
	\alpha_{T} (t) = \max_{a \in \lbrace \mathcal{E} \rbrace} \left( \frac{ \alpha_{a, T} (t) }{ \max\limits_{\text{all time}}  \alpha_{a, T} (t) } \right).
\end{equation}
Given a set of kinetics data $ \lbrace \mathcal{D} \rbrace $ and target species $ \lbrace \mathcal{T} \rbrace $, the overall importance of species \emph{S} to the target species set is
\begin{equation}	\label{E:R_S}
	\overline{ R_S } = \max_{ \substack{ T \in \lbrace \mathcal{T} \rbrace \\ k \in \lbrace \mathcal{D} \rbrace } } \left[ \max_{\text{all time, } k } \left( \alpha_{T} R_{T S} \right) \right].
\end{equation}

We employ the same sampling method as given by Niemeyer et al.\ \cite{Niemeyer:2010}, where constant volume autoignition simulations are performed using SENKIN \cite{Lutz:1988} with CHEMKIN-III \cite{Kee:1996}.  Species are considered unimportant and removed if their $\overline{ R_S }$ value falls below a cutoff threshold  $\varepsilon_{\text{EP}}$, which is selected using an iterative procedure based on a user-defined error limit for ignition delay prediction \cite{Niemeyer:2010}.  Reactions are eliminated if any participating species are removed.

\subsection{Graph search algorithms}

In this study we compare the results of DRGEP using basic DFS, modified DFS, basic BFS, \emph{R}-value-based BFS (RBFS), and Dijkstra's algorithm. Cormen et al.\ \cite{Cormen:2001} presented detailed discussion of and pseudocode for DFS, BFS, and Dijkstra's algorithm, while the modified DFS used by Niemeyer et al.\ \cite{Niemeyer:2010} and RBFS of Liang et al.\ \cite{Liang:2009} differ only slightly from the basic DFS and BFS.

\subsubsection{DFS-based algorithms}

The DFS initiates at a root node, in this case a target species, and explores the graph edges connecting to other nodes. The first node found is added to a last in, first out stack, then the search moves to this node and repeats.  In this manner, the search continues deeper down the graph pathway until it either reaches a node with no connections or all the connecting nodes have been explored, then backtracks one position up the stack.  The search is performed separately using each target species as the root node and the maximum OIC is stored for each species.  Lu and Law first introduced the DFS in this context for the DRG method \cite{Lu:2005}.

The modified DFS used by Niemeyer et al.\ \cite{Niemeyer:2010} for the DRGEP method (but not described in detail) differs from the basic DFS in that the OIC values for all target species are set to unity before starting the search at the first target.  The search then only repeats starting at other target species not discovered in the initial search.  The resulting OIC values combine the dependence of all targets on the species and this prevents the use of coefficient scaling based on individual target species activity.

\subsubsection{BFS-based algorithms}

The BFS initiates at the root node and explores all adjacent graph edges, adding connected nodes to a first in, first out queue.  After discovering all connected nodes, the search moves to the first node in the queue and restarts the procedure, moving to the next node in the queue after discovering all connected nodes.  Previously discovered nodes are not repeated.  As with the DFS, the search initializes at each target separately and the maximum OIC for each species is stored.

The \emph{R}-value-based BFS (RBFS), first introduced by Liang et al.\ \cite{Liang:2009,Liang:2009a} and also used by Shi et al.\ \cite{Shi:2010,Shi:2010a}, differs from the standard BFS in that PICs smaller than the error threshold $\varepsilon_{\text{EP}}$ are not explored.  In other words, the search ignores graph pathways that would lead to an OIC below the cutoff value and only allows discovery of important pathways.  This increases efficiency by avoiding exploration of unimportant pathways but causes the RBFS to depend on the value of $\varepsilon_{\text{EP}}$, unlike the other search methods.  As such, the primary application of this algorithm lies in dynamic reduction where the threshold value is known a priori, rather than general comprehensive skeletal reduction where the threshold is to be determined iteratively based on a user-defined error limit \cite{Niemeyer:2010}.

\subsubsection{Dijkstra's algorithm}

Dijkstra's algorithm, originally introduced by Dijkstra \cite{Dijkstra:1959} and discussed in detail by Cormen et al.\ \cite{Cormen:2001}, differs from DFS, BFS, and their variants because it was specifically designed to  calculate the shortest paths from a single source node to all other nodes rather than simply search the graph to find connected nodes.  As mentioned previously, calculating OIC values is a shortest path problem where the ``shortest'' path is that with the maximum product of intermediate edge weights.  This is the only modification needed to apply Dijkstra's algorithm to the current problem.  In brief, Dijkstra's algorithm functions similarly to BFS except that it starts with the set of graph nodes stored in a max-priority queue.  The algorithm finds and removes the node with the maximum OIC value (initially, the root node) from the queue and explores adjacent nodes to calculate or update OIC values.  When all neighboring nodes are explored, the procedure restarts at the node with the next-highest OIC in the queue.  The search completes when the queue is empty.  As with previously-discussed search algorithms, this is performed separately for each target species as root node.  See the \nameref{S:appendix} for pseudocode describing the algorithm as applied to the DRGEP method.

Transportation and internet routing applications routinely use Dijkstra's algorithm and so much effort has been focused on optimizing its performance.  A naive implementation requires a running time of $ O(V^2) $ where \emph{V} is the number of nodes.  For sparse graphs, i.e.\ where the number of edges \emph{E} follows $ E \sim O( V^2 / \log V )$, the runtime can be improved significantly to $ O(E \log V) $ by using an adjacency list to store the immediate neighbors of nodes and a binary max-heap priority queue \cite{Cormen:2001}. Using a Fibonacci heap can reduce the runtime further to $O(E + V \log V)$ \cite{Fredman:1987,Cormen:2001}, although in many cases binary heaps prove more efficient \cite{Wagner:2007}.  These will be compared in order to determine the most efficient implementation for DRGEP.  Briefly, a binary max-heap is a tree where each node has at most two child nodes and the associated key (in this case, OIC) is always greater than or equal to the keys of the child nodes.  By storing the nodes in this manner, a costly search is not required to find the node with the highest key during the operation of Dijkstra's algorithm.  Fibonacci heaps function similarly, but rather than a single tree, the heaps are made up of a collection of trees with more relaxed structures designed to avoid unnecessary operations.  Detailed discussion of the heap implementations is beyond the scope of this paper and we refer interested readers to Cormen et al.\ \cite{Cormen:2001}.

\section{Results and discussion}
\label{S:results}

\subsection{Reliability of graph search algorithms}

In order to demonstrate the dependence of DFS- and BFS-based methods on the order of species in a reaction mechanism, all five graph search algorithms are used with DRGEP to generate skeletal mechanisms from the detailed mechanism for \emph{n}-heptane of Curran et al.\ \cite{Curran:1998}, containing 561 species and 2539 reactions, where the order of species is randomly shuffled.  Basic DFS, basic BFS, RBFS, and Dijkstra's algorithm are used both with and without the target-based coefficient scaling while the modified DFS is not compatible with this and so is used without it.  Autoignition chemical kinetics data are sampled from initial conditions at 1000 K, 1 atm, and equivalence ratios of 0.5--1.5.  Oxygen, nitrogen, \emph{n}-heptane, and the hydrogen radical are selected as target species.

\begin{figure}[htbp]
	\centering
	\includegraphics[width=\linewidth]{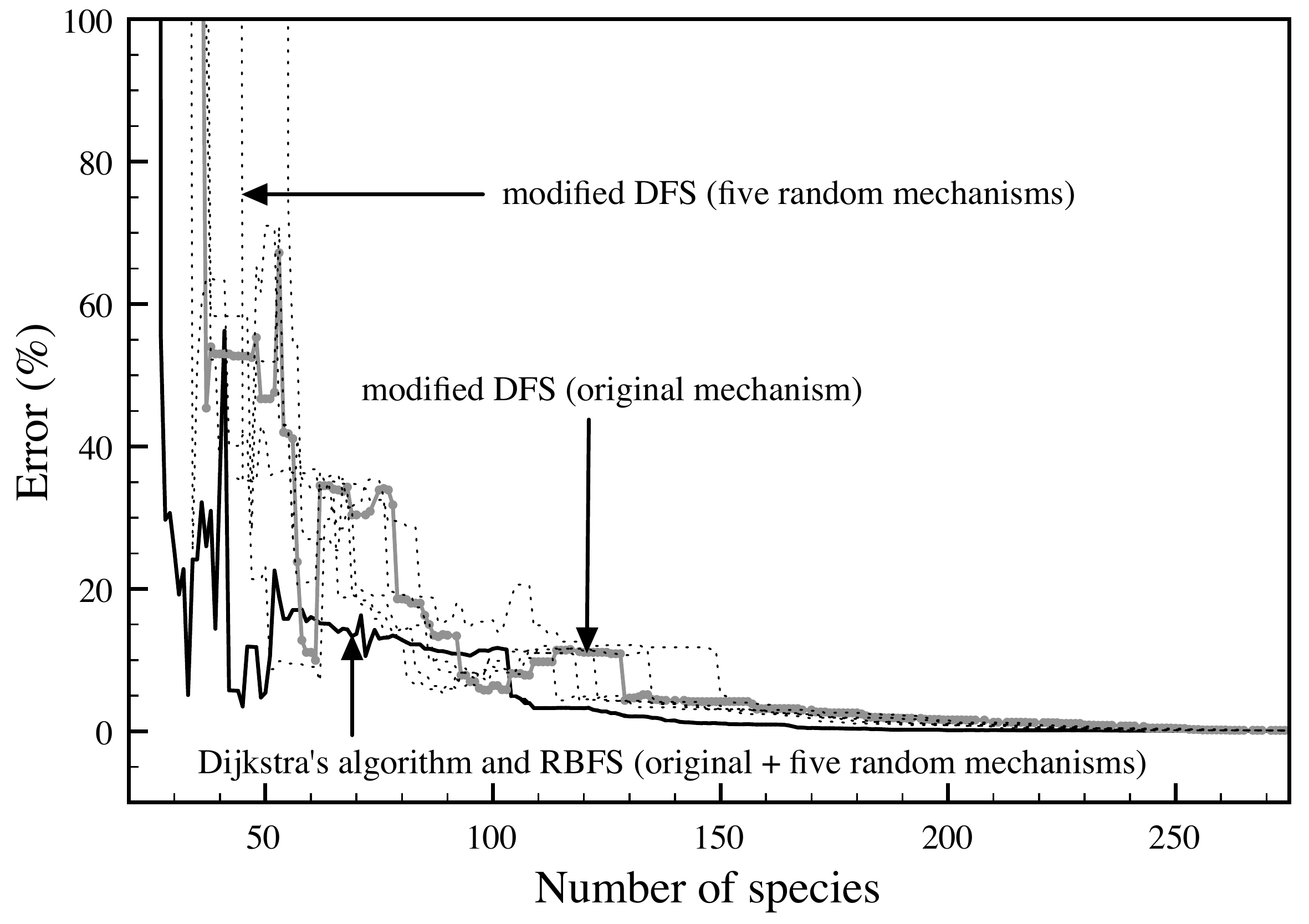}
	\caption{Error in autoignition delay prediction as a function of number of species for skeletal mechanisms of \emph{n}-heptane generated by DRGEP using Dijkstra's algorithm and RBFS with coefficient scaling and modified DFS.  The solid lines indicate the results using the original detailed mechanism while the dashed lines indicate the results obtained from five versions with randomly-shuffled species.  All 12 results using Dijkstra's algorithm and RBFS coincide.}
	\label{F:nhept_1}
\end{figure}

Figure \ref{F:nhept_1} shows the ignition delay error of skeletal mechanisms at varying levels of detail generated by DRGEP using RBFS and Dijkstra's algorithm with coefficient scaling and the modified DFS.  Basic DFS and BFS demonstrate similar dependence on species order to the modified DFS and thus are omitted from Fig.\ \ref{F:nhept_1} for clarity, but the skeletal mechanism sizes from each may be different as will be shown.  Similarly, the coefficient scaling comparison is also omitted because its application does not affect dependence on species order, although the resulting skeletal mechanism sizes are different.  Comparison of results from the original mechanism and five mechanisms with randomly shuffled species illustrates the dependence of the modified DFS on the order of species in the mechanism while RBFS and Dijkstra's algorithm produce consistent results regardless of species order.  This is because DFS and BFS explore graphs based on the order of nodes while Dijkstra's algorithm follows the strongest path independent of order \cite{Cormen:2001}.  In addition, Fig.\ \ref{F:nhept_1} shows that both Dijkstra's algorithm and RBFS produce smaller skeletal mechanisms before the error becomes unacceptably large.

The dependence of DFS, modified DFS, and BFS on species order stems from the fact that these algorithms do not in general calculate the correct OIC for every species and as such can underestimate the importance of species, causing unwarranted removal.  RBFS produces the same results as Dijkstra's algorithm because it prevents exploration of unimportant paths, i.e.\ paths with PICs (see Eq.\ \ref{E:rABp}) less than the threshold, and therefore only discovers paths that lead to an OIC greater than the threshold value.  While this value may be smaller than the correct OIC as calculated by Dijkstra's algorithm, the species is nonetheless considered important and retained in the skeletal mechanism.  This subtle error could be significant, though, if the DRGEP method is followed by a further reduction stage such as sensitivity analysis that depends on the OIC values \cite{Niemeyer:2010}.

\subsection{Effectiveness of graph search algorithms}

Skeletal mechanisms for \emph{n}-heptane are generated covering a comprehensive range of conditions using basic DFS, modified DFS, basic BFS, and Dijkstra's algorithm.  Basic DFS, BFS, and Dijkstra's algorithm are used with and without the coefficient scaling in order to determine its effect as well.  Autoignition chemical kinetics data are sampled from initial conditions at 600--1600 K, 1--20 atm, and equivalence ratios of 0.5--1.5.  Oxygen, nitrogen, \emph{n}-heptane, and the hydrogen radical are selected as target species, and the error limit in ignition delay prediction is 30\%.  RBFS is not compared here because it is not suited for a comprehensive skeletal reduction due to its dependence on threshold value; additionally, based on the results in the previous section we assume the resulting skeletal mechanism would match that from Dijkstra's algorithm.

The results using the original mechanism of Curran et al.\ \cite{Curran:1998} are summarized in Table \ref{T:n-hep1}.  All search methods lead to skeletal mechanisms with similar performance, but Dijkstra's algorithm with coefficient scaling produces the most compact skeletal mechanism.  BFS and modified DFS produce similar results while basic DFS is unable to generate a comparable skeletal mechanism for the given error limit.  This weakness is most likely due to the fact that basic DFS finds only the first path from target to species, which typically contains many intermediate species and severely underestimates the OICs for some important species.  The modified DFS performs better by inserting the other targets into the pathways and therefore increasing the OIC values for important species, while BFS finds the path with the shortest number of intermediate nodes \cite{Cormen:2001}.

Table \ref{T:n-hep1} also demonstrates the need for coefficient scaling.  Both basic DFS and Dijkstra's algorithm generate more compact skeletal mechanisms with the scaling, while BFS actually produces a slightly larger skeletal mechanism.  Without scaling, Dijkstra's algorithm is unable to produce a more compact skeletal mechanism than the other methods, despite its ability to generate results independent of species order in the parent mechanism.

\subsection{Efficiency of graph search algorithms}

% one millisec resolution of CPU timer, need to use particularly large mechanism so that speed of all algorithms and speedups adequately resolved
% system used for analysis: compute node (hgar) with 2 Xeon dual-core 3.0 GHz CPUs and 8GB memory
The computational costs of DFS, modified DFS, basic BFS, RBFS, and Dijkstra's algorithm are compared by applying the DRGEP method to the detailed mechanism of Westbrook et al.\ \cite{Westbrook:2009} for \emph{n}-alkanes covering \emph{n}-octane through \emph{n}-hexadecane, which contains 2115 species and 8157 reactions.  Kinetics data are sampled from constant volume autoignition of \emph{n}-decane with initial conditions covering 600--1600 K, 1--20 atm, and equivalence ratios of 0.5--1.5.  The efficiency improvements available to Dijkstra's algorithm including use of adjacency list, binary heap, and Fibonacci heap are also compared to the naive implementation.

\begin{figure}[htbp]
	\centering
	\includegraphics[width=\linewidth]{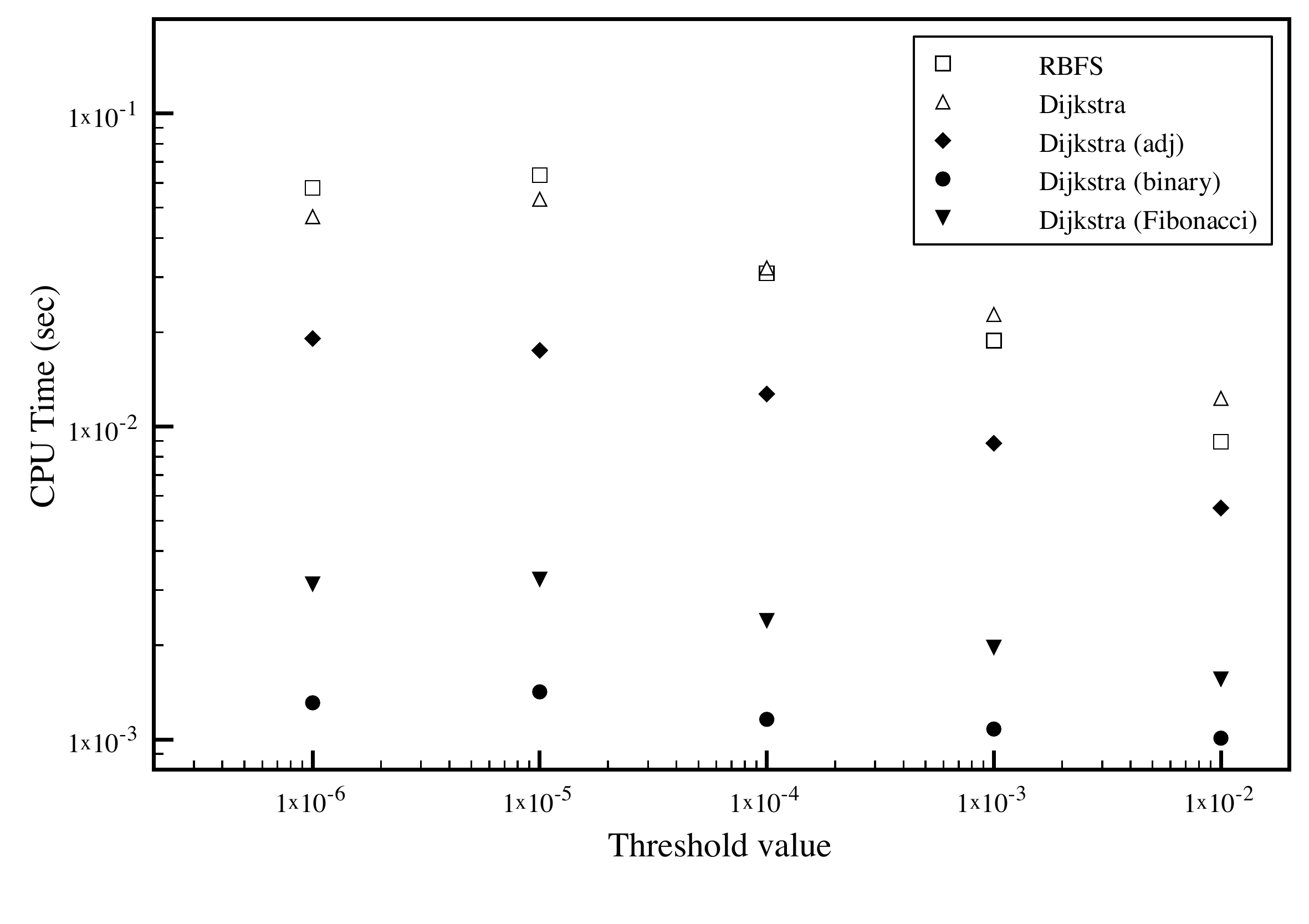}
	\caption{Comparison of average CPU time cost as a function of threshold value using RBFS and Dijkstra's algorithm, which includes the naive implementation, use of adjacency list, binary heaps, and Fibonacci heaps. The kinetics data used are generated from \emph{n}-decane autoignition over a range of initial temperatures and pressures, and at varying equivalence ratios, from a detailed mechanism consisting of 2115 species and 8157 reactions \cite{Westbrook:2009}.}
	\label{F:cost_RBFS}
\end{figure}

The average computational costs of DFS, modified DFS, BFS, Dijkstra's algorithm, and its improvements are listed in Table \ref{T:cost_comp}.  In order to perform a fair comparison, a version of Dijkstra's algorithm modified to similarly depend on the threshold value is compared with RBFS.  Figure \ref{F:cost_RBFS} shows the average computational cost of RBFS, Dijkstra's algorithm, and its improvements as a function of threshold value.  The most important result shown in Table \ref{T:cost_comp} and Fig.\ \ref{F:cost_RBFS} is that Dijkstra's algorithm implemented with binary heaps is faster than all other search algorithms, and is in fact more efficient than the same algorithm implemented with Fibonacci heaps even though the theoretical time limit with binary heaps is higher.  This result is consistent with literature comparisons for sparse graphs \cite{Cherkassky:1996,Goldberg:1996}.

\section{Concluding remarks}
\label{S:conclusion}

Graph search algorithms used in the DRGEP method for skeletal mechanism reduction are compared.  Basic DFS, basic BFS, modified DFS, RBFS, and Dijkstra's algorithm are implemented in DRGEP and used to generate (1) skeletal mechanisms covering a limited range of conditions using a randomly shuffled detailed mechanism for \emph{n}-heptane to determine the dependence of results on species order and (2) skeletal mechanisms covering a comprehensive range of conditions to determine the effectiveness of each algorithm.  The RBFS algorithm is not used to generate a comprehensive skeletal mechanism because it is not suitable for use when the cutoff threshold is not known a priori.  Both Dijkstra's algorithm and RBFS are able to generate consistent results independent of species order, and Dijkstra's algorithm used with target-based coefficient scaling generates a more compact skeletal mechanism than the other methods.  Even with the improved search algorithm, however, the size of the resulting skeletal mechanism (131 species and 651 reactions) is larger than that of DRGEP with sensitivity analysis (108 species and 406 reactions) \cite{Niemeyer:2010}, suggesting that the post-DRGEP sensitivity analysis is still required.

%efficiency
In addition to the reliability and effectiveness of the graph search algorithms, the efficiency is also compared due to its importance to dynamic DRGEP approaches.  Efficiency improvements available to Dijkstra's algorithm are also implemented and compared to the other search algorithms.  Dijkstra's algorithm with a binary heap priority queue runs two orders of magnitude faster than the other search methods and is the most efficient implementation of Dijkstra's algorithm.  As such, this approach is recommended for use in dynamic skeletal reduction using DRGEP.  Dynamic approaches that avoid entirely repeating the search by updating graph edge values have also been developed \cite{Roditty:2004,Wagner:2007} and will be the focus of future work.

\section*{Acknowledgments}
This work was supported by the National Science Foundation under Grant No.\ 0932559 and the Department of Defense through the National Defense Science and Engineering Graduate Fellowship program.

\section*{Appendix}
\label{S:appendix}

The following pseudocode describes Dijkstra's algorithm for use in the DRGEP method to calculate the OICs for all species relative to a target species \emph{T} using the set of DIC values $r_{A B}$, adapted from Cormen et al.\ \cite{Cormen:2001}.  The adjacency list \emph{adj} contains the list of adjacent (directly connected) species in the directed graph.  Operations are left in general form to be applicable to other implementations of Dijkstra's algorithm using heaps as well as the basic version.  The procedure MAKEQ (\emph{Q}) generates the max-priority queue \emph{Q}, MAXQ (\emph{R}, \emph{Q}) returns the node with the highest OIC value, REMQ (\emph{u}, \emph{Q}) removes the node \emph{u} from \emph{Q}, and UPDATE (\emph{R}, \emph{v}, $R_{\text{tmp}}$) increases the OIC value for node \emph{v} to $R_{\text{tmp}}$.

\noindent
\textbf{subroutine} Dijkstra ( $r_{A B}$, \emph{adj}, \emph{T} )
\begin{algorithmic}
	\STATE $R \leftarrow \text{zeros}$
	\STATE $R(T) \leftarrow 1.0$
	\STATE \textbf{call} MAKEQ (\emph{Q})
	\WHILE{ $ Q \ne \emptyset $ }
		\STATE $u \leftarrow $ MAXQ (\emph{R}, \emph{Q})
		\STATE \textbf{call} REMQ (\emph{u}, \emph{Q})
		\FOR{ each node $v \in adj(u)$ }
			\STATE $R_{\text{tmp}} \leftarrow R(u) \cdot r_{u v}$
			\IF{ $R_{\text{tmp}} > R(v)$ } \STATE \textbf{call} UPDATE (\emph{R}, \emph{v}, $R_{\text{tmp}}$) \ENDIF
		\ENDFOR
	\ENDWHILE
	\STATE \textbf{return} $R$
\end{algorithmic}
\textbf{end subroutine} Dijkstra

\bibliography{refs.bib}

\begin{thebibliography}{26}
\expandafter\ifx\csname natexlab\endcsname\relax\def\natexlab#1{#1}\fi
\providecommand{\bibinfo}[2]{#2}

\bibitem[{Lu and Law(2005)}]{Lu:2005}
\bibinfo{author}{T.~Lu}, \bibinfo{author}{C.~K. Law}, \bibinfo{journal}{Proc.
  Combust. Inst.} \bibinfo{volume}{30} (\bibinfo{year}{2005})
  \bibinfo{pages}{1333--1341}.

\bibitem[{Lu and Law(2006{\natexlab{a}})}]{Lu:2006}
\bibinfo{author}{T.~Lu}, \bibinfo{author}{C.~K. Law},
  \bibinfo{journal}{Combust. Flame} \bibinfo{volume}{146}
  (\bibinfo{year}{2006}{\natexlab{a}}) \bibinfo{pages}{472--483}.

\bibitem[{Lu and Law(2006{\natexlab{b}})}]{Lu:2006a}
\bibinfo{author}{T.~Lu}, \bibinfo{author}{C.~K. Law},
  \bibinfo{journal}{Combust. Flame} \bibinfo{volume}{144}
  (\bibinfo{year}{2006}{\natexlab{b}}) \bibinfo{pages}{24--36}.

\bibitem[{Lu and Law(2008)}]{Lu:2008}
\bibinfo{author}{T.~Lu}, \bibinfo{author}{C.~K. Law},
  \bibinfo{journal}{Combust. Flame} \bibinfo{volume}{154}
  (\bibinfo{year}{2008}) \bibinfo{pages}{153--163}.

\bibitem[{Zheng et~al.(2007)Zheng, Lu, and Law}]{Zheng:2007}
\bibinfo{author}{X.~L. Zheng}, \bibinfo{author}{T.~Lu}, \bibinfo{author}{C.~K.
  Law}, \bibinfo{journal}{Proc. Combust. Inst.} \bibinfo{volume}{31}
  (\bibinfo{year}{2007}) \bibinfo{pages}{367--375}.

\bibitem[{Pepiot-Desjardins and Pitsch(2008)}]{Pepiot-Desjardins:2008}
\bibinfo{author}{P.~Pepiot-Desjardins}, \bibinfo{author}{H.~Pitsch},
  \bibinfo{journal}{Combust. Flame} \bibinfo{volume}{154}
  (\bibinfo{year}{2008}) \bibinfo{pages}{67--81}.

\bibitem[{Sun et~al.(2010)Sun, Chen, Gou, and Ju}]{Sun:2010}
\bibinfo{author}{W.~Sun}, \bibinfo{author}{Z.~Chen}, \bibinfo{author}{X.~Gou},
  \bibinfo{author}{Y.~Ju}, \bibinfo{journal}{Combust. Flame}
  \bibinfo{volume}{157} (\bibinfo{year}{2010}) \bibinfo{pages}{1298--1307}.

\bibitem[{Niemeyer et~al.(2010)Niemeyer, Sung, and Raju}]{Niemeyer:2010}
\bibinfo{author}{K.~E. Niemeyer}, \bibinfo{author}{C.~J. Sung},
  \bibinfo{author}{M.~P. Raju}, \bibinfo{journal}{Combust. Flame}
  \bibinfo{volume}{157} (\bibinfo{year}{2010}) \bibinfo{pages}{1760--1770}.

\bibitem[{Zs{\'e}ly et~al.(2009)Zs{\'e}ly, Nagy, Simmie, and
  Curran}]{Zsely:2009}
\bibinfo{author}{I.~G. Zs{\'e}ly}, \bibinfo{author}{T.~Nagy},
  \bibinfo{author}{J.~M. Simmie}, \bibinfo{author}{H.~J. Curran}, in:
  \bibinfo{booktitle}{4th European Combustion Meeting},
  \bibinfo{number}{810045, 2009}.

\bibitem[{Liang et~al.(2009{\natexlab{a}})Liang, Stevens, and
  Farrell}]{Liang:2009}
\bibinfo{author}{L.~Liang}, \bibinfo{author}{J.~Stevens},
  \bibinfo{author}{J.~T. Farrell}, \bibinfo{journal}{Proc. Combust. Inst.}
  \bibinfo{volume}{32} (\bibinfo{year}{2009}{\natexlab{a}})
  \bibinfo{pages}{527--534}.

\bibitem[{Liang et~al.(2009{\natexlab{b}})Liang, Stevens, Raman, and
  Farrell}]{Liang:2009a}
\bibinfo{author}{L.~Liang}, \bibinfo{author}{J.~G. Stevens},
  \bibinfo{author}{S.~Raman}, \bibinfo{author}{J.~T. Farrell},
  \bibinfo{journal}{Combust. Flame} \bibinfo{volume}{156}
  (\bibinfo{year}{2009}{\natexlab{b}}) \bibinfo{pages}{1493--1502}.

\bibitem[{Shi et~al.(2010{\natexlab{a}})Shi, Liang, Ge, and Reitz}]{Shi:2010}
\bibinfo{author}{Y.~Shi}, \bibinfo{author}{L.~Liang}, \bibinfo{author}{H.-W.
  Ge}, \bibinfo{author}{R.~D. Reitz}, \bibinfo{journal}{Combust. Theor. Model.}
  \bibinfo{volume}{14} (\bibinfo{year}{2010}{\natexlab{a}})
  \bibinfo{pages}{69--89}.

\bibitem[{Shi et~al.(2010{\natexlab{b}})Shi, Ge, Brakora, and
  Reitz}]{Shi:2010a}
\bibinfo{author}{Y.~Shi}, \bibinfo{author}{H.-W. Ge}, \bibinfo{author}{J.~L.
  Brakora}, \bibinfo{author}{R.~D. Reitz}, \bibinfo{journal}{Energy Fuels}
  \bibinfo{volume}{24} (\bibinfo{year}{2010}{\natexlab{b}})
  \bibinfo{pages}{1646--1654}.

\bibitem[{Lu and Law(2009)}]{Lu:2009}
\bibinfo{author}{T.~Lu}, \bibinfo{author}{C.~K. Law}, \bibinfo{journal}{Prog.
  Energy Comb. Sci.} \bibinfo{volume}{35} (\bibinfo{year}{2009})
  \bibinfo{pages}{192--215}.

\bibitem[{Herbinet et~al.(2010)Herbinet, Pitz, and Westbrook}]{Herbinet:2010}
\bibinfo{author}{O.~Herbinet}, \bibinfo{author}{W.~J. Pitz},
  \bibinfo{author}{C.~K. Westbrook}, \bibinfo{journal}{Combust. Flame}
  \bibinfo{volume}{157} (\bibinfo{year}{2010}) \bibinfo{pages}{893--908}.

\bibitem[{Cormen et~al.(2001)Cormen, Leiserson, Rivest, and
  Stein}]{Cormen:2001}
\bibinfo{author}{T.~H. Cormen}, \bibinfo{author}{C.~E. Leiserson},
  \bibinfo{author}{R.~L. Rivest}, \bibinfo{author}{C.~Stein},
  \bibinfo{title}{Introduction to Algorithms}, MIT Press, Cambridge, MA,
  \bibinfo{edition}{2nd} edition, \bibinfo{year}{2001}.

\bibitem[{Dijkstra(1959)}]{Dijkstra:1959}
\bibinfo{author}{E.~W. Dijkstra}, \bibinfo{journal}{Numer. Math.}
  \bibinfo{volume}{1} (\bibinfo{year}{1959}) \bibinfo{pages}{269--271}.

\bibitem[{Curran et~al.(1998)Curran, Gaffuri, Pitz, and
  Westbrook}]{Curran:1998}
\bibinfo{author}{H.~Curran}, \bibinfo{author}{P.~Gaffuri},
  \bibinfo{author}{W.~Pitz}, \bibinfo{author}{C.~K. Westbrook},
  \bibinfo{journal}{Combust. Flame} \bibinfo{volume}{114}
  (\bibinfo{year}{1998}) \bibinfo{pages}{149--177}.

\bibitem[{Westbrook et~al.(2009)Westbrook, Pitz, Herbinet, Curran, and
  Silke}]{Westbrook:2009}
\bibinfo{author}{C.~K. Westbrook}, \bibinfo{author}{W.~J. Pitz},
  \bibinfo{author}{O.~Herbinet}, \bibinfo{author}{H.~J. Curran},
  \bibinfo{author}{E.~J. Silke}, \bibinfo{journal}{Combust. Flame}
  \bibinfo{volume}{156} (\bibinfo{year}{2009}) \bibinfo{pages}{181--199}.

\bibitem[{Lutz et~al.(1988)Lutz, Kee, and Miller}]{Lutz:1988}
\bibinfo{author}{A.~E. Lutz}, \bibinfo{author}{R.~J. Kee},
  \bibinfo{author}{J.~A. Miller}, \bibinfo{title}{{SENKIN}: A {FORTRAN} program
  for predicting homogeneous gas phase chemical kinetics with sensitivity
  analysis}, \bibinfo{howpublished}{Sandia National Laboratories Report No.
  {SAND} 87-8248}, \bibinfo{year}{1988}.

\bibitem[{Kee et~al.(1996)Kee, Rupley, Meeks, and Miller}]{Kee:1996}
\bibinfo{author}{R.~J. Kee}, \bibinfo{author}{F.~M. Rupley},
  \bibinfo{author}{E.~Meeks}, \bibinfo{author}{J.~A. Miller},
  \bibinfo{title}{{CHEMKIN-III}: A {FORTRAN} chemical kinetics package for the
  analysis of gas-phase chemical and plasma kinetics},
  \bibinfo{howpublished}{Sandia National Laboratories Report No. {SAND}
  96-8216}, \bibinfo{year}{1996}.

\bibitem[{Fredman and Tarjan(1987)}]{Fredman:1987}
\bibinfo{author}{M.~L. Fredman}, \bibinfo{author}{R.~E. Tarjan},
  \bibinfo{journal}{J. ACM} \bibinfo{volume}{34} (\bibinfo{year}{1987})
  \bibinfo{pages}{596--615}.

\bibitem[{Wagner and Willhalm(2007)}]{Wagner:2007}
\bibinfo{author}{D.~Wagner}, \bibinfo{author}{T.~Willhalm},
  \bibinfo{journal}{Lect. Notes Comput. Sci.} \bibinfo{volume}{4393}
  (\bibinfo{year}{2007}) \bibinfo{pages}{23--36}.

\bibitem[{Cherkassky et~al.(1996)Cherkassky, Goldberg, and
  Radzik}]{Cherkassky:1996}
\bibinfo{author}{B.~Cherkassky}, \bibinfo{author}{A.~Goldberg},
  \bibinfo{author}{T.~Radzik}, \bibinfo{journal}{Math. Program.}
  \bibinfo{volume}{73} (\bibinfo{year}{1996}) \bibinfo{pages}{129--174}.

\bibitem[{Goldberg and Tarjan(1996)}]{Goldberg:1996}
\bibinfo{author}{A.~V. Goldberg}, \bibinfo{author}{R.~E. Tarjan},
  \bibinfo{title}{Expected performance of {Dijkstra's} shortest path
  algorithm}, \bibinfo{type}{Technical Report} \bibinfo{number}{TR-530-96},
  {NEC} Research Institute, Princeton University, \bibinfo{year}{1996}.

\bibitem[{Roditty and Zwick(2004)}]{Roditty:2004}
\bibinfo{author}{L.~Roditty}, \bibinfo{author}{U.~Zwick},
  \bibinfo{journal}{Lect. Notes Comput. Sci.} \bibinfo{volume}{3221}
  (\bibinfo{year}{2004}) \bibinfo{pages}{580--591}.

\end{thebibliography}
\bibliographystyle{elsarticle-num-CNF.bst}

\pagebreak

%tables

% table 1: n-heptane results
\begin{table}[htbp]
\begin{center}
\begin{tabular}{l l c c c}
\toprule
\multicolumn{2}{c}{Algorithm} & \# Species & \# Reactions & Max.\ Error \\ \midrule
DFS		& no scaling	& 461	& 2304	& 19\% 	\\
 		& scaling		& 449 	& 2267 	& 19\% 	\\ \midrule
mod.\ DFS & 	& 173 	& 868 	& 28\% 	\\ \midrule
BFS 		& no scaling	& 180	& 891	& 29\% 	\\
		& scaling		& 207 	& 921 	& 25\% 	\\ \midrule
Dijkstra 	& no scaling	& 178	& 986	& 23\% 	\\
		& scaling		& 131 	& 651 	& 17\%	\\ \bottomrule
\end{tabular}
\caption{Comparison of \emph{n}-heptane skeletal mechanism sizes generated by the DRGEP method using DFS, BFS, and Dijkstra's algorithm with and without coefficient scaling and modified DFS.  The original mechanism contains 561 species and 2539 reactions \cite{Curran:1998}.}
\label{T:n-hep1}
\end{center}
\end{table}

% table 2: algorithm speed comparison
\begin{table}[htbp]
\begin{center}
\begin{tabular}{c c c}
\toprule
Algorithm & Mean cost (sec) & Cost normalized by Dijkstra \\ \midrule
DFS			& 1.48 \e{-1}	& 2.20 	\\ \midrule
mod.\ DFS 	& 1.46 \e{-1} 	& 2.18 	\\ \midrule
BFS			& 8.30 \e{-2} 	& 1.24 	\\ \midrule
Dijkstra (naive)	& 6.72 \e{-2} 	& 1.00 	\\ 
Dijkstra (adj)	& 1.18 \e{-2}	& 0.176	\\
Dijkstra (binary heap)	& 1.5 \e{-3}	& 0.028 \\
Dijkstra (Fibonacci heap)	& 3.4 \e{-3}	& 0.051 \\
\bottomrule
\end{tabular}
\caption{Comparison of average CPU time costs using DFS, modified DFS, BFS, and Dijkstra's algorithm, which includes the naive implementation, use of adjacency list, binary heaps, and Fibonacci heaps. The kinetics data used are generated from \emph{n}-decane autoignition over a range of initial temperatures and pressures, and at varying equivalence ratios, from a detailed mechanism consisting of 2115 species and 8157 reactions \cite{Westbrook:2009}.}
\label{T:cost_comp}
\end{center}
\end{table}

\pagebreak
\pagebreak

%list of captions for figures

\listoffigures

\pagebreak

%figure: n-heptane Dijkstra vs DFS

%figure: RBFS and Dijkstra's cost comparison (depends on threshold value)

\end{document}